\documentstyle[aps,epsfig]{revtex}
\def\bbox#1{\mbox{\boldmath $#1$}}

\begin{document}

\draft

\title{Excitation of $S_{11}$ resonances in pion scattering and
pion photoproduction on the proton}
\author{  Guan-Yeu Chen, Sabit Kamalov\cite{Sabit} and Shin  Nan Yang}
\address{Department of Physics, National Taiwan University, Taipei, Taiwan
10764, Republic of China}
\author{Dieter Drechsel and Lothar Tiator}
\address{Institut f\"ur Kernphysik, Universit\"at Mainz, 55099 Mainz, Germany}

\date{\today}
\maketitle

\begin{abstract}
A self-consistent analysis of  pion scattering and pion
photoproduction  within a coupled channels dynamical model is
presented.
In the case of pion photoproduction, we obtain background
contributions to the imaginary part of the $S$-wave multipole
which differ considerably from the result based on the K-matrix
approximation. Within the dynamical model these background
contributions become large and negative in the region of the
$S_{11}(1535)$ resonance. Due to this fact much larger resonance
contributions are required in order to explain the results of the
recent multipole analyses. For the first $S_{11}(1535)$ resonance
we obtain as a value of the dressed electromagnetic helicity
amplitude: $A_{1/2}=(72\pm 2) \times 10^{-3}$GeV$^{-1/2}$. Similar
values can be derived from eta photoproduction if one takes the
same total width ($\Gamma_R=95\pm 5$ MeV)  as in pion scattering
and pion photoproduction. The combined analysis yields
considerable strength at invariant mass $W\ge1750$ MeV, which can
be explained by a third and a fourth $S_{11}$ resonance with the
masses $1846\pm 47$ and $2113\pm 70$ MeV.
\end{abstract}

\vspace{0.3cm}

\pacs{PACS numbers: 13.60.Le, 13.75.Gx, 14.20.Gk, 25.20.Lj}

\section{Introduction}

Hadron spectroscopy has always played an important role in
unraveling the underlying quark dynamics. For example, the pattern
of similarity between the eight baryons and the eight pseudoscalar
mesons led to the epochal discovery of SU(3) symmetry.

There are  44 nonstrange baryon states listed in the Particle Data
Group \cite{PDG2002}, 22 in $T=\frac 12$ and another 22 in $T=\frac 32$
channels. Among them, 18 are rated four-star and 6
rated three-star. The rest are weakly excited states with at most
fair evidence of existence. Even though the existence of the
four-star baryon resonances is certain, some very large
discrepancies exist in their properties as obtained from different
analyses. One example is the extracted width of the four-star
$S_{11}(1535)$ state given as $66$ MeV \cite{Arndt95}, $120\pm20$
MeV \cite{Hoehler79}, $151\pm27$ MeV \cite{Manley92}, $ 151-198$
MeV \cite{Feuster98}, and $270\pm50$ MeV \cite{Cutkovsky79}. The
differences between different analyses arise mostly from the data
set included in the analysis or the separation of background and
resonance contributions. This model dependence in
the extraction of the resonance properties has made it difficult
to test the predictions of theoretical models with existing data.

At present the resonance properties are extracted mainly from $\pi
N$ scattering, 2$\pi$ and $\eta$ production and pion
photoproduction using different approaches (for details see
Refs.~\cite{BT91,Batinic,Vrana,Pit}). The first coupled channel
analysis that combines pion and eta data was done in
Ref.~\cite{BT91} within the isobar model. Later, more sophisticated
models were developed which account for background
contributions. Most of them are based on the  solution of
coupled-channels equations by use of the K-matrix approximation,
i.e., by ignoring off-shell intermediate scattering states. A
motivation for such an approximation is the result of
Ref.~\cite{Oset98}, where it was shown that off-shell effects
could be incorporated into the vertex renormalization
of the lowest order Lagrangians.

On the other hand, in the analysis of pion scattering and pion
photoproduction within dynamical
models~\cite{Tanabe,Yang85,NBL}, the off-shell dynamics (i.e., the
dynamics at short distance) is taken into account. Within this
framework we have recently developed a meson-exchange (MEX) model
for pion-nucleon scattering~\cite{hung}  which gives good
agreement with the data up to 400 MeV pion lab energy. In
addition, we have also constructed a dynamical model for pion
electromagnetic production \cite{KY99,DMT} which describes
the $\pi^0$ photo- and electro-production data near threshold
\cite{kamalov01} and most of the existing data up to the second resonance region.

In this paper, we extend our meson-exchange $\pi N$ model in the
$S_{11}$ channel up to $2$ GeV by explicitly introducing a set of
$S_{11}$ resonances into the model. The results are then fed into the pion photoproduction
model to analyze the existing $_pE_{0^+}$ multipole. The $S_{11}$
channel is of interest for several reasons.  First of all, the
first resonance $S_{11}(1535)$, which lies only 48 MeV above the
$\eta N$ threshold, has a remarkably large $\eta N$ branching
ratio. This necessitates the inclusion of the $\eta N$ channel into our
MEX $\pi N$ model. Secondly, the analyses based solely on pion
photoproduction always underestimate the $A^p_{1/2}$ helicity
amplitude of $S_{11}(1535)$ with a value around $60 \times
10^{-3}$GeV$^{-1/2}$, while extractions from the $(\gamma,\eta)$
data give a value close to and above  $100 \times
10^{-3}$GeV$^{-1/2}$ \cite{PDG2002}. Lastly, there have been
suggestions \cite{Giannini02,Saghai02} that there could exist a
third $S_{11}$ resonance in the neighborhood of $1800-1900$
MeV, in addition to the well-known resonances at 1535 and 1650
MeV. A consistent analysis of both $\pi N$ scattering and pion
photoproduction reactions can shed new light on the mentioned issues
concerning helicity amplitudes and higher
resonance as well as reduce the large uncertainties in the width
obtained from previous analyses.

\section{$\pi N$ scattering}

Our MEX  $\pi N$ model results by use of a three-dimensional
reduction scheme of the Bethe-Salpeter equation for a model
Lagrangian involving $\pi, N, \Delta, \rho,$ and $\sigma$ fields.
Details  can be found in Ref.\cite{hung}. Here we only present the
general scheme to extend the model to the case of coupled $\pi$,
$\eta$, and $2\pi$ channels, including in addition the couplings
with baryon resonances in the $S_{11}$ partial wave. To do this,
we first enlarge the Hilbert space to include a bare $S_{11}$
particle $R$ which acquires a width by couplings with $\pi N$ and
$\eta N$ channels via the Lagrangian
\begin{eqnarray}
{\cal{L_I}}= ig^{(0)}_{\pi NR}\bar R\bbox{\tau}N\cdot\bbox{\pi} +
ig^{(0)}_{\eta NR}\bar R N\eta + h.c., \label{lagr}
\end{eqnarray}
where $N, R, \bbox{\pi},$ and $\eta$ denote the field operators
for the nucleon, bare $R$, pion and eta, respectively. The
full $t$-matrix can then be written as a system of coupled
equations,
\begin{eqnarray}
t_{ij}(E)= v_{ij}(E)+\sum_k  v_{ik}(E)\,g_k(E)\, t_{kj}(E)\,,
\label{t_ij}
\end{eqnarray}
where $i$ and $j$ denote the $\pi$ or $\eta$ channel and $E=W$ is
the total center of mass energy. Equation~(\ref{t_ij}) is a system of
three-dimensional coupled integral equations which is derived from
the four-dimensional Bethe-Salpeter equation using a
three-dimensional reduction scheme with a
relativistic propagator $g_k$ for the free $kN$ system ($k=\pi$,
or $\eta$), constructed according to the Cooper-Jennings reduction
scheme~\cite{CJ89}.

In general, the potential $ v_{ij}$ is a sum of non-resonant
$(v^B_{ij})$ and  bare resonance $(v^R_{ij})$ terms,
\begin{eqnarray}
v_{ij}(E)=  v^B_{ij}(E)+ v^R_{ij}(E)\,. \label{v_ij}
\end{eqnarray}
The non-resonant term $v^B_{\pi\pi}$ for the $\pi N$ elastic
channels contains contributions from the $s$- and $u$-channels, Born
terms and $t$-channel contributions with $\omega$, $\rho$, and
$\sigma$ exchange. In the present work the parameters in
$v^B_{\pi\pi}$ are fixed from the analysis of the pion scattering
phase shifts for the $s-$ and $p-$waves at low energies ($W<1300$
MeV). In channels involving the eta, $v^B_{i\eta}$ is taken to be
zero because of the small $\eta NN$ coupling~\cite{TBK}. The
bare resonance contribution $v^R_{ij}(E)$ arises from
excitation and de-excitation of the resonance $R$ via the
interaction Lagrangian of Eq. (\ref{lagr}).

Let us first consider the case with  only one $S_{11}$ resonance
contributing in the energy region of
interest, i.e., $W < 2$ GeV. The corresponding potential
$v^R_{ij}(E)$ can then be symbolically expressed in the form of
\begin{eqnarray}
v^R_{ij}(q,q';E)=\frac{f_i(\tilde\Lambda_i,q;E)\,g_i^{(0)}\,g_j^{(0)}\,
f_j(\tilde\Lambda_j,q';E)}{E-M_R^{(0)}+
i\frac{1}{2}\Gamma^R_{2\pi}(E)} \,, \label{v_R1}
\end{eqnarray}
where $q$ and $q'$ are the pion (or eta) momenta in the initial
and final states, and $M_R^{(0)}$ and $g_{i(j)}^{(0)}$ are the bare
masses and resonance vertex couplings, respectively. According to
our previous $\pi N$ model, we associate with each line of the
particle $\alpha$ in a Feynman diagram the covariant form factor
$F_\alpha$
$[n_\alpha\Lambda^4_\alpha/(n_\alpha\Lambda^4_\alpha+(p^2_\alpha
-m^2_\alpha)^2)]^{n_\alpha}$, where $p_\alpha$, $m_\alpha$, and
$\Lambda_\alpha$ are the four-momentum, mass, and cut-off
parameter of particle $\alpha$, respectively. The quantity $f_i$
of Eq.~(\ref{v_R1}) is then a product of three form
factors $F_{\alpha}$, each of them corresponding to one of the three legs
associated with the considered vertex. As a result, $f_i$ depends
on the  product of three cut-off parameters, i.e., $\tilde\Lambda_\pi \equiv
(\Lambda_N,\Lambda_R,\Lambda_\pi)$. We refer the readers to Refs.
\cite{hung,PJ91} for details and only mention that $n_\alpha=10$
is used in the present work.

In Eq. (\ref{v_R1}) we have added a phenomenological term
$\Gamma^R_{2\pi}(E)$ in the resonance propagator in order to take
into account the $\pi\pi N$ decay
channel. Therefore, our resonance propagator is not purely "bare"
but includes renormalization (or "dressing") effects due to the
coupling with the $\pi\pi N$ channel. Using this prescription we
assume that the additional non-resonant mechanisms of coupling
with the $\pi\pi N$ channel are small. Following
Refs.\cite{Lvov,MAID98} we take $\Gamma^R_{2\pi}(E)$  as
\begin{eqnarray}
\Gamma^R_{2\pi}(E)=\Gamma^{0,R}_{2\pi}\left(\frac{q_{2\pi}}{q_0}\right)^{2l+4}
\left(\frac{X^2+q^2_0}{X^2+q^2_{2\pi}}\right)^{l+2}\,,
\label{G_2pi}
\end{eqnarray}
where $l$ is the pion orbital momentum, $q_{2\pi}$ the momentum
of the compound $2\pi$ system with mass $2m_{\pi}$ and
$q_0=q_{2\pi}$ at $E=M_{R}^{(0)}$. We note that this form takes
account of the correct energy behavior of the phase space near the
three-body threshold~\cite{Lvov}. The parameter $X$ is fixed at
500 MeV as suggested in Ref.~\cite{MAID98}, and the quantity
$\Gamma^{0,R}_{2\pi}$ is associated with the $2\pi$ decay width at
resonance. In general, $\Gamma_{2\pi}^{0,R}$ can  be considered as a
free parameter. However, its value is strongly correlated with the
values of the unknown coupling constants $g_i^{(0)}$. In particular, a
small $g_i^{(0)}$ usually requires a large $\Gamma^{0,R}_{2\pi}$.
Therefore, we will fix the value for $\Gamma^{0,R}_{2\pi}$ by use
of recent knowledge about the decay modes of the resonances and
their total Breit-Wigner widths~\cite{PDG2002}.

Summarizing our parametrization of the potential $v^R_{ij}$ in the
case of  coupled pion and eta channels, we would like to emphasize
that, in general, one isolated resonance contains five free
parameters, the bare mass $M^{(0)}_R$, the decay width
$\Gamma^0_{2\pi}$, two bare coupling constants $g^{(0)}_i$ and
$g^{(0)}_j$, and one cutoff parameter $\Lambda_R$.

In the channel of interest, $S_{11}$, there are two well-known
four-star resonance states, $S_{11}(1535)$ and $S_{11}(1650)$, and
one one-star resonance, $S_{11}(2090)$. In the Hypercentral
Constituent Quark Model~\cite{Giannini02}, a third and fourth
$S_{11}$ resonance with energies 1860 and 2008~MeV were predicted.
The generalization of our coupled channels model for multiple
resonances with the same quantum numbers is simply
\begin{eqnarray}
v^R_{ij}(q,q';E)=\sum_{n=1}^{N} v^{R_n}_{ij}(q,q';E), \label{v_RN}
\end{eqnarray}
with additional parameters for the bare masses, widths, coupling
constants and cut-off parameters for each  resonance.

We first start  with the analysis of Re $t_{ij}$ and Im $t_{ij}$
for pion scattering and eta production in the energy range
$1100~$MeV $<W<1750$ MeV where the $S_{11}(1535)$ and $S_{11}(1650)$
resonances are very pronounced. The results of our best fit of
$t_{\pi\pi}$ in this energy range with only these two resonances
included are shown in Fig. 1 by the dotted curves. We are not able
to improve our results in the region $W>1800$ MeV  without
additional $S_{11}$ resonances. Next we extend the energy range
up to $W=2000$ MeV and add a third resonance with
the parameters for the first resonance fixed as obtained above. Our
results for this case are shown by the dash-dotted curves, which
correspond to a bare mass of the third $S_{11}$ resonance
$M_3^{(0)}=1901$ MeV.  We find that  this value is very stable and
changes only within 2\% if the energy range is increased up to
2200 MeV. However, this does not remove the remaining discrepancy,
in particular for the imaginary part at $W>2000$ MeV. We find that
the only way to improve the agreement with the data in this energy
range is to introduce a fourth resonance. Our final fit results
($\chi^2/120=4.56$) with four $S_{11}$ resonances are
shown by the solid lines in Fig. 1. The obtained value for the
bare mass of the fourth $S_{11}$ resonance is $M_4^{(0)}=2160$
MeV. Note that in Fig. 1 the background contributions (dashed
curves) are defined by the equation
\begin{eqnarray}
\tilde t_{\pi\pi}^{B}(E)= v_{\pi\pi}^{B}(E)+
v_{\pi\pi}^{B}(E)\,g(E)\,\tilde t_{\pi\pi}^B(E)\ ,
\label{NR_back}
\end{eqnarray}
which are hereafter called the "nonresonant background", i.e., the background
with nonresonant rescattering. In the following Fig.~2 we show our results
for the t-matrix of the $\pi N\rightarrow \eta N$ reaction in the
$S_{11}$ channel, which clearly indicate the presence  of the
$\eta$ decay mode in the second and third $S_{11}$ resonance regions.

Now let us  turn to the more sophisticated part of the analysis,
namely, the extraction of the physical (or "dressed") masses,
partial widths and branching ratios of the resonances.  As was
pointed out in Ref.~\cite{Vrana}, the procedure is certainly model
dependent.  This is mainly connected with the question  how to
separate background and resonance contributions in a well-defined
way. The solution to this problem becomes more difficult with an
increasing number of overlapping resonances in the same
channel. Below we   present our solution to this problem.

First, we determine the physical mass using as
definition for the  contribution of the $n$-th  resonance $R_n$ in
the elastic pion scattering channel:
\begin{eqnarray}
 t^{R_n}_{\pi\pi}(E)=v^{R_n}_{\pi\pi}(E)+\sum_k
v^{R_n}_{\pi k}(E)\, g_{k}(E)\,t_{k\pi}(E)\,,
\label{T_Rn}
\end{eqnarray}
with $ t_{k\pi}(E)$  the full $t$-matrix obtained from the
solution
  of Eq. (\ref{t_ij}). It is easy to  see that
Eq. (\ref{T_Rn}) corresponds to the following decomposition of the
full $t_{\pi\pi}$ matrix:
\begin{eqnarray}
 t_{\pi\pi}(E)= t^{B}_{\pi\pi}(E)+
\sum_{n=1}^{N}  t^{R_n}_{\pi\pi}(E)\,,
\label{T_pipi}
\end{eqnarray}
where the new background operator is defined as
\begin{eqnarray}
t_{\pi\pi}^{B}(E)= v_{\pi\pi}^{B}(E)+ \sum_k v_{\pi k
}^{B}(E)\,g_k(E)\, t_{k \pi}(E)\,.
\label{R_back}
\end{eqnarray}
 We call this the "resonant background", because it contains resonance
 contributions via the full scattering matrix $t_{k \pi}(E)$.

It can be shown that the sum of the two terms on the r.h.s. of Eq.
(\ref{T_Rn}) can be expressed by a single term as represented by
the diagram on the l.h.s. in Fig. 3, which  consists of a bare
initial $\pi N R_n$ vertex, followed by a dressed resonance
propagator and then finally decay through a dressed $\pi N R_n$
vertex. It now becomes obvious that $t^{R_n}_{\pi\pi}$ should have
the standard Breit-Wigner form, $\sim
(W-M_{R_n}+i\Gamma_{R_n}/2)^{-1}$, near the resonance position.
The energy $W$ where Re $t^{R_n}_{\pi\pi}$ crosses zero can be
considered as the physical mass of the resonance $R_n$. The total
width can be determined by the full width at half maximum of
$Im\,t^{R_n}_{\pi\pi}(W)$.
The branching ratios for the pion, eta and two-pion decay modes
can be extracted from the corresponding self-energies calculated
by projection of the $t^{R_n}_{\pi\pi}$-matrix on the standard
Breit-Wigner form.

Our final results for the resonance parameters are summarized in
Table I. Here we also compare with the results of the
GW-Giessen collaboration, Pitt-ANL collaboration and Kent State
University (KSU) group, taken from Ref.~\cite{Pit}. Our parameters
for the first and second $S_{11}$ resonances are close to the
Pitt-ANL results. Both models predict a total width
$\Gamma_{R_1}\simeq$100 MeV for the first and
$\Gamma_{R_2}\simeq$200 MeV for the second $S_{11}$ resonance. Note that the
value obtained for the total width of the $S_{11}(1535)$
is very close to the recent result of Ref.~\cite{Oset02} (94~MeV)
and Ref.~\cite{SAID} (106 MeV). However, there is no
agreement with the results of the GW-Giessen model, which yields
$\Gamma_{R_1}\simeq$230 MeV, and the KSU model, which yields
$\Gamma_{R_2}\simeq$108 MeV. In the last case the reason for the
discrepancy could be
the absence of the $\eta N$ production data in the fit of that
reference. Within our
model we can get a similar result for $\Gamma_{R_2}$ if we also
 exclude this
channel from our data base. Further differences lie in the
parameters for the third $S_{11}$. The position and total width of
this resonance in our and the Pitt-ANL analyses are close, but our
model suggests a strong one-pion decay mode (about 40\%), while
the Pitt-ANL value is only 17\%. The situation with the strength
of the eta decay mode is opposite: we find a branching ratio of
12\% as opposed to 41\% in the Pitt-ANL analysis.  As we will see
below such differences have visible consequences for the pion and
eta photoproduction. We further note that in the case of the third
$S_{11}$, results similar to ours were also obtained in
Ref.~\cite{Batinic} ($\Gamma_{\pi}/\Gamma_R=51$\%).

\section{Pion photoproduction}

The above analysis of elastic $\pi N$
scattering indicates the existence of four $S_{11}$ resonances.
Let us now check this result by an independent analysis of pion
photoproduction using the dynamical model developed in Refs.
\cite{Yang85,KY99,DMT}, hereafter called the DMT (Dubna-Mainz-Taipei)
model. Concerning the details of the DMT model, we refer the reader
to Ref.~\cite{KY99}.

The t-matrix  for pion photoproduction in the dynamical model is
\begin{eqnarray}
t_{\gamma\pi}(E)=v_{\gamma\pi}+\sum_k v_{\gamma
k}\,g_k(E)\,t_{k\pi}(E), \label{tgampi1}
\end{eqnarray}
with $v_{\gamma k}$ the transition potential for the $\gamma N
\rightarrow k N$ reaction ($k=\pi$ or $\eta$), $t_{k \pi}$ the
full $k N$ scattering t-matrix of Eq. (\ref{t_ij}), and $g_k$
the free $k N$ propagator.

If the transition potential $v_{\gamma\pi}$
consists of two terms,
\begin{eqnarray}
v_{\gamma\pi}(E)=v_{\gamma\pi}^B + v_{\gamma\pi}^{R}(E),
\label{vgampi}
\end{eqnarray}
where $v_{\gamma\pi}^B$ is the background transition potential and
$v_{\gamma\pi}^{R}(E)$ the contribution of a bare resonance R, we
may decompose the resulting t-matrix into two terms \cite{KY99},
\begin{eqnarray}
t_{\gamma\pi}(E)=t_{\gamma\pi}^B(E) + t_{\gamma\pi}^{R}(E),
\label{tgampi2}
\end{eqnarray}
where
\begin{eqnarray}
t_{\gamma\pi}^B(E)=v_{\gamma\pi}^B+\sum_k v_{\gamma k}^B\,g_k(E)
\,t_{k\pi}(E),\label{t BRa} \\
t_{\gamma\pi}^R(E)=v_{\gamma\pi}^R+\sum_k v_{\gamma k}^R\,g_k(E)
\,t_{k\pi}(E). \label{t_BR}
\end{eqnarray}
In our numerical calculations we have neglected the contribution
of the  $\eta$ channel in the intermediate states in Eq. (\ref{t
BRa}), because this contribution  is found to be much smaller than
for the $\pi$ channels. We further note that all the processes which start
with the excitation of the resonance by the bare $\gamma N R$
vertex are summed up in $t_{\gamma\pi}^R$. This is similar to our
definition Eq. (\ref{T_Rn}) of the resonance contribution for $\pi
N$ scattering.  Using the decomposition of Eqs.
(\ref{tgampi2}-\ref{t_BR}) we can now extract the value of the
bare $\gamma N R$ vertex.  As in the case of pion scattering (see
Eq. (\ref{R_back})), the corresponding background
$t^B_{\gamma\pi}$ is called "resonant background" since it
contains the full pion scattering $t$-matrix. Note that, for
example, in Ref. \cite{Sato} the background is defined
differently,
\begin{eqnarray}
\tilde
t_{\gamma\pi}^B(E)=v_{\gamma\pi}^B+v_{\gamma\pi}^B\,g_{\pi}(E)\,
\tilde t^B_{\pi\pi}(E)\,, \label{t_back}
\end{eqnarray}
where $\tilde t^B_{\pi\pi}$ is defined by Eq. (\ref{NR_back}).
This definition corresponds to the so-called nonresonant (smooth)
background, because it contains none of the resonance
contributions. The corresponding resonance term $\tilde
t^R_{\gamma\pi}=t_{\gamma\pi}-\tilde t^B_{\gamma\pi}$ describes a
resonance with a dressed $\gamma N R$ vertex.

It can be easily proved that the following relation holds between
$t^R_{\gamma\pi}$ and $\tilde {t}^R_{\gamma\pi}$:
\begin{eqnarray}
\tilde t_{\gamma\pi}^R(E)=t_{\gamma\pi}^R(E)
+v_{\gamma\pi}^B\,g_{\pi}(E)\,{\tilde t}^R_{\pi\pi}(E),
\label{bare_dressed}
\end{eqnarray}
where ${\tilde t}^R_{\pi\pi}(E)=t_{\pi\pi}(E)-{\tilde
t}^B_{\pi\pi}(E)$ is the dressed resonance contribution in pion
scattering, as is graphically illustrated in Fig. 4.

The background potential $v_{\gamma\pi}^B$  contains Born terms
with an energy dependent mixing of pseudovector and pseudoscalar
$\pi NN$ coupling and t-channel vector meson exchanges
\cite{MAID98}. The mixing parameters and coupling constants were
determined from an analysis of the nonresonant multipoles. The
standard physical multipoles in a channel $\alpha=\{l,j,I\}$ can
then be expressed as
\begin{eqnarray}
  t^{B,\alpha}_{\gamma\pi}(q_E,k)
&= &v^{B,\alpha}_{\gamma\pi}(q_E,k)[1+iq_E
F_{\pi\pi}^{(\alpha)}(q_E,q_E;E)]
 - \frac {P}{\pi}\int_0^{\infty}\frac{q'^2 dq'}{{\cal
M}(q')}\frac{ F_{\pi\pi}^{(\alpha)}(q_E,q';E)\,
v^{B,\alpha}_{\gamma\pi}(q',k)} {E -E_{\pi N}(q') }, \label{B_DMT}
\end{eqnarray}
where $F^{(\alpha)}_{\pi\pi}$ is the pion-scattering amplitude
with the on-shell value $F^{(\alpha)}_{\pi\pi}(q_E,q_E)=[\eta_{\alpha}
\exp(2i\delta_{\alpha})-1]/2iq_E,$ with $\delta_{\alpha}$ the
phase shift and $\eta_{\alpha}$  the inelasticity parameter, and
${\cal M}(q)=E_{\pi}(q)E_N(q)/E_{\pi N}(q)$ the relativistic
pion-nucleon reduced mass. We mention in passing that the
so-called "K-matrix" approximation, as in the case of
MAID~\cite{MAID98,sabit1} and many others models, neglects the
principal value integral in Eq.~(\ref{B_DMT}), and parametrizes
the background in terms of on-shell
pion rescattering only. In this case, the off-shell rescattering associated
with the principal value integral contribution is
phenomenologically absorbed in the resonance parameters, while in
the DMT model it is considered as a part of the background.
Therefore, the resonance parts in DMT and K-matrix approach are
different: in the  DMT model the resonance is described by the
amplitude $t_{\gamma\pi}^R$ with a bare electromagnetic vertex
and, as we will see below,  in the models based on the K-matrix
approximation, the resonance description is essentially given by
$\tilde t_{\gamma\pi}^R$ with a dressed electromagnetic vertex.

Following Ref.~\cite{MAID98}, we assume a Breit-Wigner form for
the resonance contribution $t_{\gamma\pi}^{R,\alpha}(W)$,
\begin{equation}
t_{\gamma\pi}^{R,\alpha}(W)\,=\,{\bar{\cal
A}}_{\alpha}^R\, \frac{f_{\gamma R}(W)\Gamma_R\,M_R\,f_{\pi
R}(W)}{M_R^2-W^2-iM_R\Gamma_R}\,,
\label{BW}
\end{equation}
where $f_{\pi R}$ is the usual Breit-Wigner factor describing the
decay of a resonance $R$ with total width $\Gamma_{R}(W)$ and
physical mass $M_R$. The expressions for $f_{\gamma R}, \, f_{\pi
R}$ and $\Gamma_R$ are given in Ref.~\cite{MAID98}.  In the DMT
model the electromagnetic form factor ${\bar{\cal A}}_{\alpha}^R$
describes the bare $\gamma NR$ vertex.  This is a free parameter
to be determined from the experimental data.

In Fig. 5 (upper panel) we see that the resonant background  in
the DMT model (dash-dotted curve) is very important, in particular
for $ W> 1450$ MeV where it becomes large and negative. This is in
contrast to the prediction based on the K-matrix approximation
(dashed curve). The difference comes mainly from the principal
value integral contribution in Eq. (\ref{B_DMT}). Such a
background will thus require a much stronger resonance
contribution  in order to describe the results of the recent
partial wave analysis of Ref.\cite{SAID}. Consequently, the
dynamical model predicts much larger values for the
electromagnetic form factors (or helicity amplitudes $A_{1/2}$)
than those obtained with the K-matrix approximation.

In order to estimate the new values for the resonance parameters,
we will first fit  Im~$_pE_{0+}$ in the photon energy range $1075
< W < 2300$ MeV only, thereby assuming that
$v^{R,\alpha}_{\gamma\pi}=\sum_{n=1}^{4}
v^{R_n,\alpha}_{\gamma\pi}$. The results of our fit are presented
in Tables II-IV and Fig. 5. We would like to stress that, since the
DMT background is large and negative even at $ W>1770$ MeV,   the
best fit requires two new $S_{11}$ resonances with masses 1800 MeV
and 2042 MeV, in addition to the well known resonances
$S_{11}(1535)$ and $S_{11}(1650)$. In fact the $\chi^2$ of the fit
improves from 64 to 3.5 by introducing these two additional
resonances. This result clearly indicates that, in agreement with
our previous findings for pion scattering, our pion
photoproduction model calls for a low-lying third $S_{11}$
resonance  which may be one of the missing resonances predicted by
quark models~\cite{Giannini02}.

Our next important result concerns the value of the helicity
amplitude for the first $S_{11}(1535)$ resonance. Here we expect
to get more reliable information  by analyzing the observables
(differential cross sections, beam, target and recoil asymmetries)
in the full energy range $1075<W<1770$ MeV, similar to the work of
Ref.\cite{sabit1}. In other words, we have to perform a new
partial wave analysis including the new background description.
For this purpose the resonances in the higher partial waves have
to be taken into account as well. The details of our fitting
procedure are described in Ref. \cite{sabit1}. We only note that
our analysis includes experimental data  for $E_{\gamma}<1200$
MeV.  In the proton channel the corresponding data base contains
14880 data points. We fix the parameters for the third and fourth
$S_{11}$ resonances using the results for
 Im $_pE_{0+}$ as described above. Our final results are shown in
Fig. 6 and summarized in Table V. In general they are consistent
with the earlier results obtained  directly from Im $_pE_{0+}$.

Our final value for the bare helicity amplitude of the first
$S_{11}(1535)$ resonance is $A_{1/2}({\rm bare})=116\pm 3 \times
10^{-3}$GeV$^{-1/2}$. In order to obtain the dressed value,  we first
have to determine the nonresonant background $\tilde
t_{\gamma\pi}^{B\alpha}$, as given by Eq. (\ref{t_back}).  The
result for this background, shown in Fig.~7 (left panel) by the
dashed curve,  is small near resonance position. The dressed value
for the $A_{1/2}$ can be calculated directly from Im
$_pE_{0+}^{(1/2)}$ using the relation~\cite{Arndt90}
\begin{equation}
A_{1/2}= -\sqrt{\frac{2\pi M_R \Gamma_R q_R}{k_R m_N
\beta_{\pi}}}\,{\rm Im}\, _pE_{0+}^{(1/2)}\,C_{\pi N}\,,
\label{A_dressed}
\end{equation}
where $q_R$ and $k_R$ are the pion and photon momentum,
respectively, at $W=M_R=1528$ MeV and $C_{\pi N}=-\sqrt{3}$ is an
isospin factor .  Subtracting  the contribution of the other
$S_{11}$ resonances we obtain Im $_pE_{0+}^{(1/2)} = (3.93-0.14)
\times 10^{-3}/m_{\pi+}$, where the last number is the
contribution of the nonresonant background $\tilde
t_{\gamma\pi}^{B\alpha}$, and $A_{1/2}({\rm dressed})=(72\pm
2)\times 10^{-3}$GeV$^{-1/2}$. Another separation of the resonance
and background contributions can be obtained by use of the
K-matrix approximation for pion rescattering. In this case Im
$_pE_{0+}^{(1/2)} = (3.93 - 0.42) \times 10^{-3}/m_{\pi+}$ and the
corresponding helicity amplitude  $A_{1/2}$(K-matrix)$=(67\pm
2)\times 10^{-3}$GeV$^{-1/2}$, which is very close to the dressed
value obtained above.

As a last step, we extract the bare and dressed values for
$A_{1/2}$ from eta photoproduction. The formalism for this
reaction is similar to the pion photoproduction case, the only
difference being that in the eta channel it is important to take
 account of the coupling to the pion channel. The resonant
background $t_{\gamma\eta}^B$ can then be written as
\begin{eqnarray}
t_{\gamma\eta}^B(E)=v_{\gamma\eta}^B+v_{\gamma\eta}^B\,
g_{\eta}(E)\,t_{\eta\eta}(E)
+v_{\gamma\pi}^B\,g_{\pi}(E)\,t_{\pi\eta}(E)\,, \label{t_Beta}
\end{eqnarray}
where $t_{\eta\eta}(E)$ and $t_{\pi\eta}(E)$ are the full
t-matrices describing eta scattering and eta production by pions,
respectively, as  obtained by solving Eq. (\ref{t_ij}).

Our results for the resonant background in the $_pE_{0^+}$ channel
are shown in Fig. 7 (right panel) by the dash-dotted curve. We can
see that this background is about 30\% of the total amplitude
(solid curves), and that it originates mainly from the coupling to
the pion channel. Using Eq. (\ref{A_dressed}) (with the eta
momentum for $q_R$, branching ratio $\beta_{\eta}=0.5,$ and
$C_{\eta N}=-1$) we can now extract $A_{1/2}$(bare),
$A_{1/2}$(dressed) and $A_{1/2}$(K-matrix). The final results and
a comparison with the results obtained for the first $S_{11}$
resonance  in pion photoproduction are given in Table VI. In
general the values of the helicity amplitudes derived from the two
reactions are consistent within 10\%. Note that our bare values
are also close to the result obtained in Ref.~\cite{SFN97}
($A_{1/2}=102 \times 10^{-3}$GeV$^{-1/2}$) by solving of the
coupled Bethe-Salpeter equations. However, our dressed value
considerably differs from the results of Ref.\cite{VTC} where
$A_{1/2}({\rm dressed})=118 \times 10^{-3}$GeV$^{-1/2}$. This is
mainly due to the different total width $\Gamma_R$, which is 191
MeV in Ref.\cite{VTC} and 95 MeV in our model. For example using
Eq.~(\ref{A_dressed}), we can easily obtain that
$A_{1/2}$(dressed)$\simeq 118 \sqrt{95/191}= 83\times
10^{-3}$GeV$^{-1/2}$ which is close to our result.

The comparison with the results of others groups are given in
Table VII. Here we present only the dressed values for the
helicity amplitude of the first $S_{11}$ resonance. Our
corresponding background is shown in Fig.~7 (left panel) by the dashed
curve which does not contain the pion loop contribution in the
electromagnetic vertex (last diagram of Fig.~4). At the resonance
position this background is very close to the backgrounds of the
Pitt-ANL and GW-Giessen groups~\cite{Pit}. Therefore, we expect
that the values for the helicity amplitudes will also be close
if the other resonance parameters ($\Gamma_R$ and $\beta_{\pi}$)
are the same. With this condition a simple recalculation of the
Pitt-ANL results yields $A_{1/2}$(dressed)$\simeq 87
\sqrt{0.35/112\cdot 95/0.40}= 74 \times 10^{-3}$GeV$^{-1/2}$,
which is close to our value. In a similar way we obtain
$A_{1/2}$(dressed)$\simeq 60 \times
10^{-3}$GeV$^{-1/2}$ for the GW-Giessen results.
Large differences remain only with the
results of the KSU analysis, and this is mainly connected with a larger
the background contribution in the KSU model (see Fig. 1 in
Ref.~\cite{Pit}).

We thus  conclude that the  large differences for the helicity
amplitude of the $S_{11}(1535)$ resonance obtained in  different
analyses, are due to the differences for the total widths and
differences in the background description. From our coupled
channel analysis of pion scattering and pion photoproduction
reactions, we obtain $\Gamma_R\simeq 100$ MeV. Note that values
close to ours were also obtained within the Pit-ANL and KSU
models~\cite{Pit} and in Refs.~\cite{Oset02,SAID}. It seems that
in order to clarify the situation for the  total width, a new
analysis of eta photoproduction is required, especially with the
additional background contributions appearing due to the coupling
to the pion channel.

\section{Conclusion}

We have performed a self-consistent analysis
of pion scattering and pion photoproduction  within a coupled
channels dynamical model.
In the case of pion photoproduction, we
obtain background contributions to the imaginary part of the $S$-
wave multipole which differ considerably from the result based on
the K-matrix approximation. Within the dynamical model these
background contributions become large and negative in the region
of the $S_{11}(1535)$ resonance. Due to this fact much larger
resonance contributions are required in order to explain the
results of the recent multipole analyses. For the first
$S_{11}(1535)$ resonance we obtain as values of the bare and
dressed electromagnetic helicity amplitudes: $A_{1/2}({\rm
bare})=(116\pm 3)\times 10^{-3}$GeV$^{-1/2}$ and $A_{1/2}({\rm
dressed})=(72\pm 2) \times 10^{-3}$GeV$^{-1/2}$. Similar values
can be derived from eta photoproduction if one takes the same
total width ($\Gamma_R=95\pm 5$ MeV)  as in pion scattering and
pion photoproduction.

At invariant energy $W\ge1750$ MeV, our analysis yields
considerable strength, which can be described by a third and a
fourth $S_{11}$ resonance with the masses $1846\pm 47$ and
$2113\pm 70$ MeV. Such resonances are also predicted by quark
models. However, our coupled channels approach yields large
resonance widths of the order of 300 MeV or more, and therefore
the energy dependence of the $S_{11}$ amplitude is not very
pronounced at these higher energies. Whether or not these
indications for higher $S_{11}$ strengths will eventually lead to
well-established resonances, will strongly depend on future
experiments and analyses of final states with uncorrelated
two-pion systems, $\rho$ and $\omega$ mesons and sequential decays
like the $\rho-\Delta$ process.

\acknowledgements

S.K. and L.T. are grateful to the Physics Department of the NTU
for the hospitality extended to them during their visits. This
work is supported in part by the National Science Council/ROC
under grant NSC  90-2112-M002-032, by the Deutsche
Forschungsgemeinschaft (SFB 443), and by a joint project NSC/DFG
TAI-113/10/0.

\newpage
\begin{table}[htbp]
\begin{center}
\begin{tabular}{|c|cccc|c|}
     & $R_1$ & $R_2$ & $R_3$ & $R_4$ \\
\hline
                 &  1520(6)  & 1677(6) &  1893(18) & 2183(21)& our result\\
                 &  1542(3)  & 1689(12)&  1822(43) & ------  & Pitt-ANL \\
$M_R$ (MeV)      &  1528(1)  & 1649(1) &  2000(1)  & ------  & KSU \\
                 &  1549(7)  & 1690(11)&  ------   & ------  & GW-Giessen\\
                 & 1520-1555 &1640-1680& $\sim$2090& ------  & PDG2002\\
\hline
                 &  95(10)   & 195(14) &  265(31)  & 427(26) & our result\\
                 & 112(19)   & 202(40) &  248(185) & ------  & Pitt-ANL\\
$\Gamma_R$ (MeV) & 111(9)    & 108(14) &  132(16)  & ------  & KSU   \\
                 & 232(19)   & 206(13) &  ------   & ------  & GW-Giessen\\
                 & 100-200   & 145-190 &  ------   & ------  & PDG2002\\
\hline
                 &  40(7)    &  75(4)  &  44(5)    & 43(2)   & our result\\
                 &  35(5)    &  74(2)  &  17(7)    & ------  & Pitt-ANL\\
$\Gamma_{\pi}/\Gamma_R$(\%)&41(3)     &  32(5)  &  10(3)    & ------  & KSU\\
                 &  30(1)    &  60(7)  &  ------   & ------  & GW-Giessen\\
                 &  35-55    &  55-90  &  ------   & ------  & PDG2002\\
\hline
                 &   40(7)   &  11(3)  &  12(7)    &  3(6)   & our result\\
                 &   51(5)   &  6(1)   &  41(4)    & ------  & Pitt-ANL\\
$\Gamma_{\eta}/\Gamma_R$ (\%)&50(4)   &  5(3)   &   3(2)    & ------  & KSU\\
                 &   63(3)   &  11(1)  &  ------   & ------  & GW-Giessen\\
                 &  30-55    &  3-10   &  ------   & ------  & PDG2002 \\
\end{tabular}
\end{center}
\caption{ $S_{11}$ resonance  parameters obtained from $\pi N$
scattering and $\pi N\rightarrow \eta N$, and comparison
with the results of the Pitt-ANL, KSU and GW-Giessen groups taken
from Ref.~\protect\cite{Pit}. Recent PDG values from
Ref.~\protect\cite{PDG2002}.}
\end{table}

\begin{table}[htbp]
\begin{center}
\begin{tabular}{|c|c|ccc|}
 Parameters  & $\beta_{\pi}=\Gamma_{\pi}/\Gamma_R$ & $M_R$  & $\Gamma_R$  & $A_{1/2}$(bare)  \\
\hline
$R_1$  &  0.40 & $1523\pm 3$ &  $97\pm 7$  &  $110\pm 7$ \\
$R_2$  &  0.75 & $1689\pm 3$ &  $164\pm 8$ &  $142\pm 6$ \\
\end{tabular}
\end{center}
\caption{ Estimation of the $S_{11}$ resonance  parameters
obtained by fitting Im $_pE_{0+}$ in the energy range 145 MeV
$<E_{\gamma}<$ 1200 MeV  (1075 MeV $< W <$ 1900 MeV), with only
two $S_{11}$ resonances. Helicity amplitudes $A_{1/2}$ are given
in units $10^{-3}$ GeV$^{-1/2}$.}
\end{table}

\begin{table}[htbp]
\begin{center}
\begin{tabular}{|c|c|ccc|}
 Parameters  & $\beta_{\pi}=\Gamma_{\pi}/\Gamma_R$ & $M_R$  & $\Gamma_R$  & $A_{1/2}$(bare)  \\
\hline
$R_1$  &  0.40 & $1523\pm 3$ &  $97\pm 7$  &  $110\pm 7$ \\
$R_2$  &  0.75 & $1689\pm 3$ &  $160\pm 11$  &  $119\pm 5$ \\
$R_3$  &  0.44 & $1962\pm 10$ & $503\pm 80$ &  $131\pm 7$ \\
\end{tabular}
\end{center}
\caption{ Estimation of $S_{11}$ resonance  parameters obtained by
fitting Im $_pE_{0+}$ in the energy range 145 MeV
$<E_{\gamma}<$ 2000 MeV  (1075 MeV $< W <$ 2300 MeV), with three
$S_{11}$ resonances ($\chi^2/95=4.7$). The parameters for the
first $S_{11}$ resonance were fixed according to Table I.
Helicity amplitudes $A_{1/2}$ are given
in units $10^{-3}$ GeV$^{-1/2}$.}
\end{table}

\begin{table}[htbp]
\begin{center}
\begin{tabular}{|c|c|ccc|}
 Parameters  & $\beta_{\pi}=\Gamma_{\pi}/\Gamma_R$ & $M_R$  & $\Gamma_R$  & $A_{1/2}$(bare)  \\
\hline
$R_1$  &  0.40 & $1523\pm 3$ (1524)  &  $97\pm 7$  &  $110\pm 7$ \\
$R_2$  &  0.75 & $1677\pm 3$ (1688)  &  $116\pm 8$  &  $83\pm 6$ \\
$R_3$  &  0.44 & $1799\pm 9$ (1861) & $314\pm 11$ &  $129\pm 9$ \\
$R_4$  &  0.43 & $2042\pm 16$ (2008) & $288\pm 29$ &  $61\pm 8$ \\
\end{tabular}
\end{center}
\caption{ Estimation of the $S_{11}$ resonance  parameters
obtained by fitting Im $_pE_{0+}$ in the energy range 145 MeV
$<E_{\gamma}<$ 2000 MeV (1075 MeV $< W <$ 2300 MeV), with four
$S_{11}$ resonances ($\chi^2/95=3.5$). The parameters for the
first $S_{11}$ resonance were fixed according to Table I. In
brackets: quark model predictions of Ref.\protect\cite{Giannini02}
for the masses $M_R$ . Helicity amplitudes $A_{1/2}$ are given
in units $10^{-3}$ GeV$^{-1/2}$.}
\end{table}

\begin{table}[htbp]
\begin{center}
\begin{tabular}{|c|ccc|}
 Parameters   & $M_R$  & $\Gamma_R$  & $A_{1/2}$(bare) \\
\hline
$R_1$ & 1528 $\pm$ 1 &  95 $\pm$ 5  &   116 $\pm$ 3 \\
$R_2$ & 1684 $\pm$ 1 &  112 $\pm$ 7  &  72  $\pm$ 2 \\
\end{tabular}
\end{center}
\caption{ Our final results for the resonance parameters of the
first two $S_{11}$ resonances obtained by fitting the observables in
the energy range 145 MeV $<E_{\gamma}<$ 1200 MeV  (1075 MeV $< W
<$ 1770 MeV). The parameters for the third and fourth $S_{11}$
resonances were fixed with the values from Table IV. The branching
ratios are $\beta_{\pi}$=0.4 and 0.75 for $R_1$ and $R_2$,
respectively.}
\end{table}

\begin{table}[htbp]
\begin{center}
\begin{tabular}{|c|c|c|}
 $A_{1/2}$  & $(\gamma , \pi)$  &  $(\gamma , \eta)$ \\
\hline
 $A_{1/2}$(bare)     & $ 116 \pm 3\quad(6.1)$  & $108 \pm 4\quad(17.8)$ \\
 $A_{1/2}$(dressed)  & $ 72\pm 2\quad(3.8)$ & $   81 \pm 3\quad(13.5)$ \\
 $A_{1/2}$(K-matrix) & $67\pm 2\quad(3.5)$  & $   83 \pm 3\quad(13.8)$ \\
\end{tabular}
\end{center}
\caption{ $A_{1/2}$(bare), $A_{1/2}$(dressed) and
$A_{1/2}$(K-matrix) helicity amplitudes
(in units $10^{-3}$ GeV$^{-1/2}$) for the $S_{11}(1535)$
resonance obtained in pion and eta photoproduction using the
coupled channel dynamical model (see text). The values correspond
to $M_R=1528$ MeV, $\Gamma_R=95$ MeV, $\beta_{\pi}=0.4$, and
$\beta_{\eta}=0.5$. In the brackets we give the imaginary part of
the corresponding $E_{0+}$ multipoles  at the resonance position
in units of $10^{-3}/m_{\pi+}$.}
\end{table}

\begin{table}[htbp]
\begin{center}
\begin{tabular}{|c|ccc|}
Ref. & $\beta_{\pi}=\frac{\Gamma_{\pi}}{\Gamma_R}$  &  $\Gamma_R$(MeV) & $A_{1/2}$(dressed)\\
\hline
 our        & $ 0.40 \pm 0.07$  & $95  \pm 5$  & $72 \pm 2$ \\
 Pitt-ANL   & $ 0.35 \pm 0.05$  & $112 \pm 19$ & $87 \pm 3$ \\
 KSU        & $ 0.41 \pm 0.03$  & $111 \pm  9$ & $42 \pm 6 $\\
 GW-Giessen & $ 0.30 \pm 0.01$  & $232 \pm 19$ & $106 \pm 5 $\\
 PDG2002    & $0.35-0.55 $       & $100-200$    & $90  \pm 30 $\\
\end{tabular}
\end{center}
\caption{ Resonance parameters and helicity amplitudes
$A_{1/2}$(dressed) (in units $10^{-3}$ GeV$^{-1/2}$) for the
$S_{11}(1535)$ resonance and comparison with the results of the
Pitt-ANL, KSU and GW-Giessen groups taken from
Ref.~\protect\cite{Pit}. Recent PDG values from
Ref.~\protect\cite{PDG2002}.}
\end{table}

\begin{figure}[htb]
\centerline{\epsfig{file=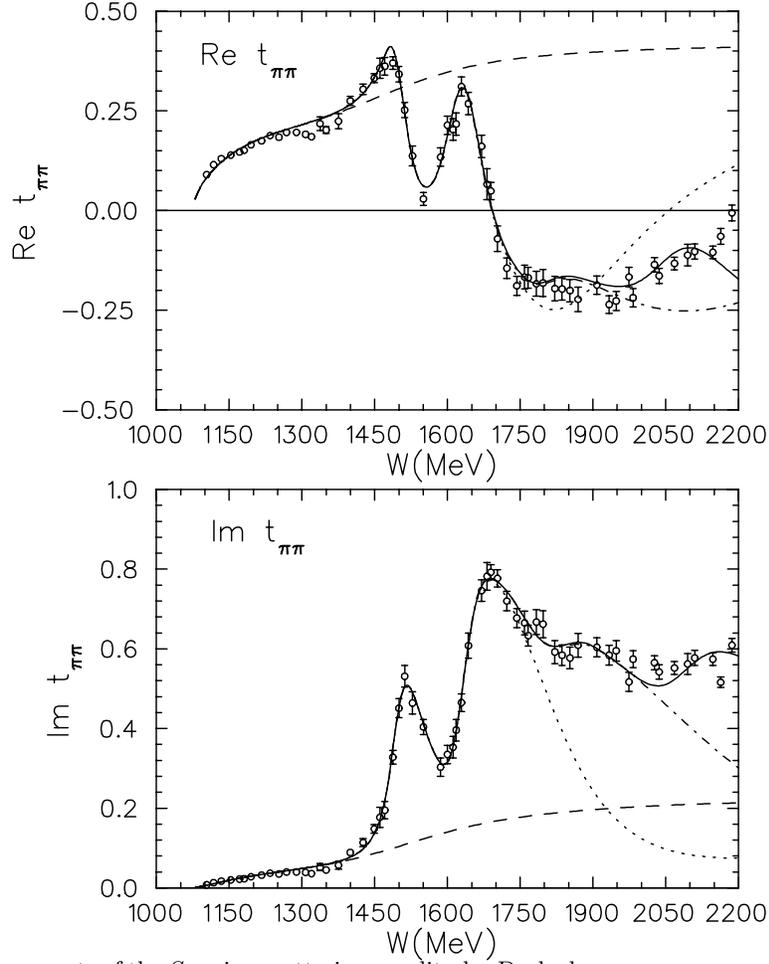,width=10.0cm, angle=0}}
\caption{Real and imaginary parts of the $S_{11}$ pion scattering
amplitude. Dashed curves: nonresonant background contribution
${\tilde t}^B_{\pi\pi}$. Dotted, dash-dotted, and solid curves:
total $t_{\pi\pi}$ amplitude obtained after the best fit with two,
three, and four $S_{11}$ resonances, respectively. Data points:
results of the single energy analysis from
Ref.\protect\cite{VPI95}.} \label{fig1}
\end{figure}

\begin{figure}[htb]
\centerline{\epsfig{file=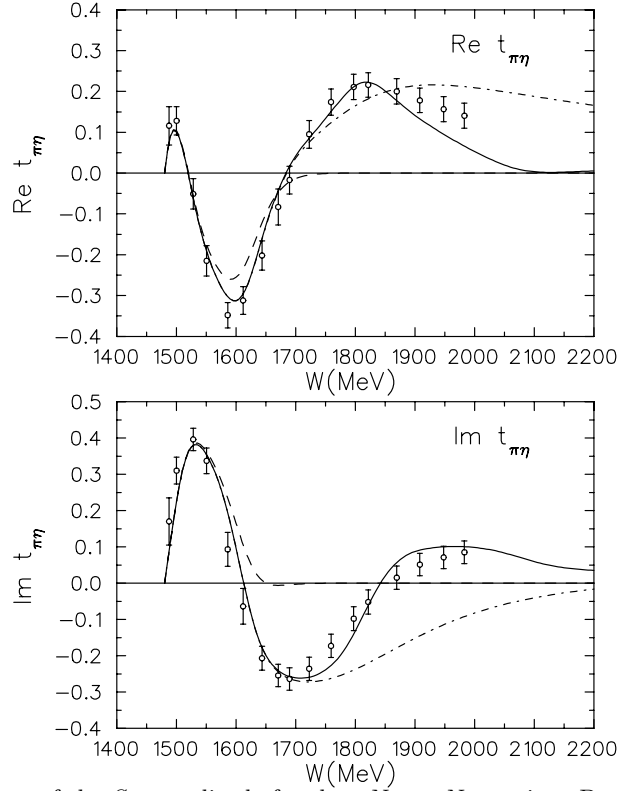,width=8.0cm, angle=0}}
\caption{Real and imaginary parts of the $S_{11}$ amplitude for
the $\pi N \rightarrow \eta N$ reaction. Dashed curves, dash-dotted,
and solid curves are the results obtained with $R_{1}$,
$R_{1}+R_{2}$, and $R_{1}+R_{2}+R_{3}$ contributions, respectively.
We did not find evidence for a fourth $S_{11}$ in this reaction. Data
points: results of the partial wave analysis from
Ref.\protect\cite{Vrana}.} \label{fig2}
\end{figure}

\begin{figure}[htb]
\centerline{\epsfig{file=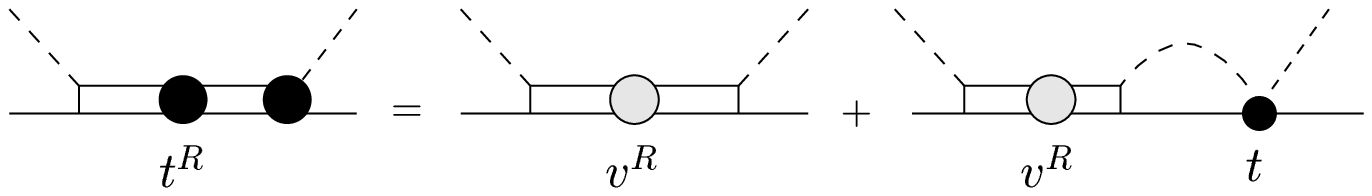,width=9.0cm, angle=0}}
\caption{Graphical representation of the resonance contribution to
pion scattering as determined by Eq. (\protect\ref{T_Rn}). The
grey circle in the resonance propagators denotes the presence of
the width associated with the $\pi\pi N$ contribution, the black
circle corresponds to the propagators with total width and
physical mass.}
\label{fig3}
\end{figure}

\begin{figure}[htb]
\centerline{\epsfig{file=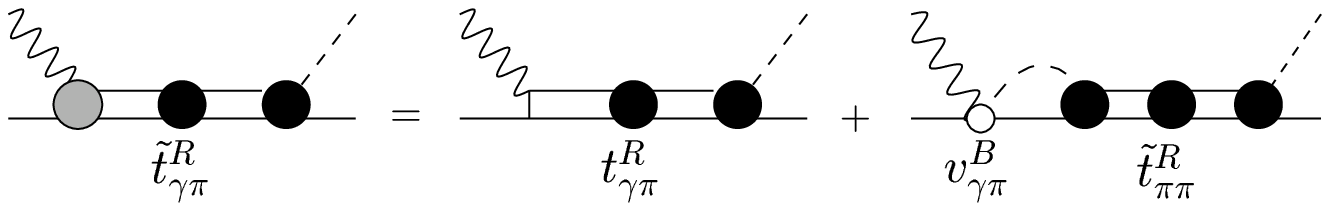,width=9.0cm, angle=0}}
\caption{Graphical representation of the resonances with dressed
and bare electromagnetic vertices.}
\label{fig4}
\end{figure}

\begin{figure}[htb]
\centerline{\epsfig{file=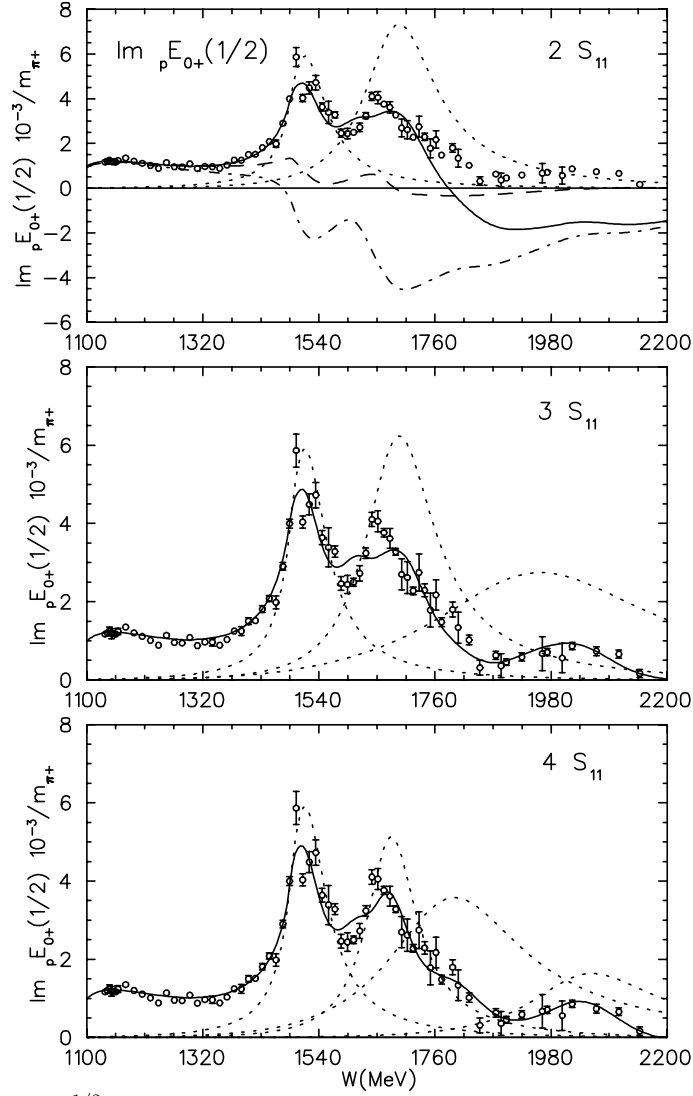,width=9.0cm, angle=0}}
\caption{Imaginary parts of the $_pE_{0+}^{1/2}$ multipoles.
Dashed and dash-dotted curves in the upper panel: background
contributions obtained using K-matrix approximation and DMT model,
respectively. Solid curves in the upper, middle and lower panels:
total multipole with two, three, and four $S_{11}$ resonances,
respectively. The individual contributions from each resonance
(with bare electromagnetic vertex) are shown by the dotted curves.
The corresponding resonance parameters are given in Tables II-IV.
Data points: results of the single energy multipole analysis from
Ref.\protect\cite{SAID}.}
\label{fig5}
\end{figure}

\begin{figure}[htb]
\centerline{\epsfig{file=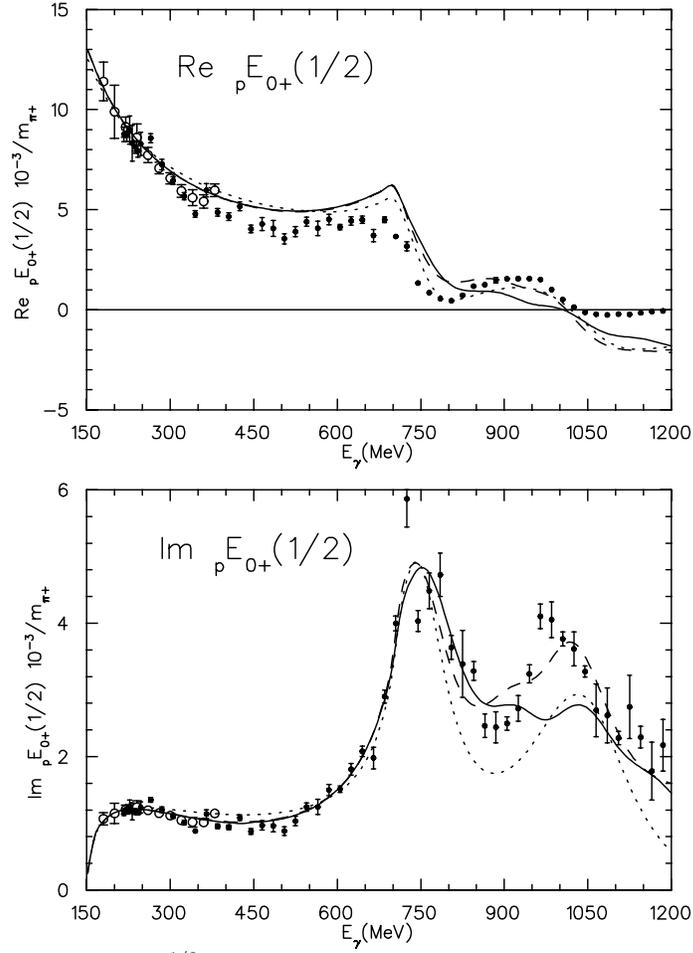,width=9.0cm, angle=0}}
\caption{Real and imaginary parts of the $_pE_{0+}^{1/2}$
multipole. Dashed and solid curves: results obtained by fitting Im
$_pE_{0+}^{1/2}$ (see Table IV) and the observables (see Table V),
respectively. Dotted curve: result of
MAID2000~\protect\cite{MAID98}. Data points: results of the single-energy
multipole analysis of Ref.\protect\cite{HDT} ($\circ$) and
Ref.\protect\cite{SAID}($\bullet$).}
\label{fig6}
\end{figure}

\newpage
\begin{figure}[htb]
\centerline{\epsfig{file=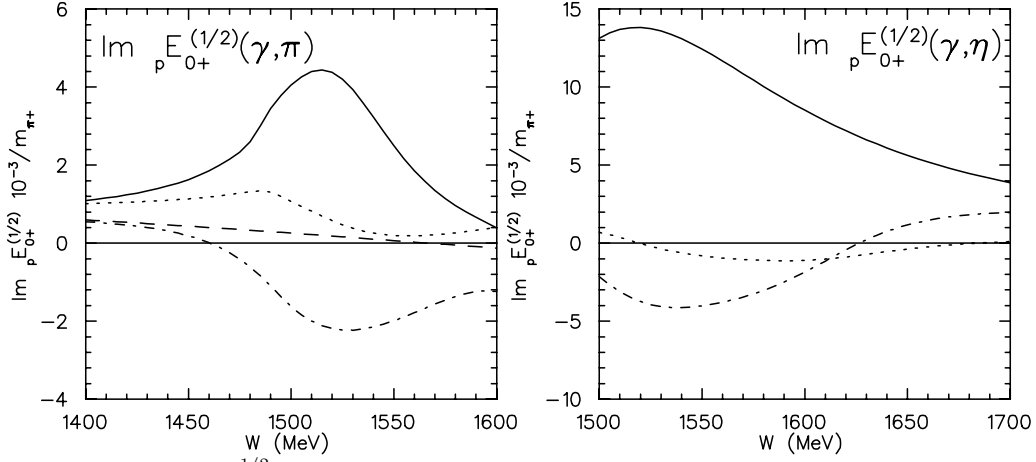,width=6.0cm, angle=90}}
\caption{Imaginary parts of the $_pE_{0+}^{1/2}$ multipoles for
pion (left figure) and eta (right figure) photoproduction on the
proton. Dash-dotted and dashed curves are resonant, $t^B$
(Eqs.(\protect\ref{tgampi2}-\protect\ref{t_BR})), and nonresonant,
$\tilde t^B$ (Eq.(\protect\ref{t_back})), backgrounds,
respectively. In the case of eta photoproduction, the main
contribution to the resonant background comes from the coupling to
the $\pi \eta$ channel, the contribution $\tilde t^B$ of the nonresonant
background vanishes. Dotted curves: background
according to the K-matrix approximation. Solid curves: total
result including contributions of $S_{11}(1535)$ resonance and
background, taken from Ref~\protect\cite{VTC}.} \label{fig7}
\end{figure}

\end{document}